\documentclass[12pt]{article}
\usepackage{authblk}
\usepackage[bookmarksnumbered, colorlinks, plainpages]{hyperref}
\usepackage{amsmath, amsthm, amscd, amsfonts, amssymb, graphicx, color, booktabs}
\numberwithin{equation}{section}
\definecolor{email}{rgb}{0.00,0.00,0.84}
\begin{document}
\setcounter{page}{1}

\title{\large \bf 12th Workshop on the CKM Unitarity Triangle\\ Santiago de Compostela, 18-22 September 2023 \\ \vspace{0.3cm}
\LARGE Flavour bounds on axions, hidden photons and sterile neutrinos}

\author{Martin Bauer \\
        Institute for Particle Physics Phenomenology, Department of Physics, 
Durham University, Durham, DH1 3LE, UK}

\date{March 30, 2024}

%\dedicatory{This paper is dedicated to Professor ABCD}
\maketitle

\begin{abstract}

Flavour-violating decays are some of the most sensitive probes for New Physics with masses below the B meson threshold.  In this talk I review the sensitivity to axion-like particles, hidden photons and heavy neutral leptons assuming minimal models where these are the only particle in addition to the Standard Model, respectively.

\end{abstract} \maketitle

\section{Introduction}
Flavour observables are usually thought of as sensitive probes of energy scales far above those that can be probed directly at particle accelerators. Heavy new particles can mediate flavour violating amplitudes at tree-level or enter loops and modify the branching ratios for flavour-violating decays or meson mixing parameters predicted by the Standard Model (SM).\footnote{See the talk by Wolfgang Altmannshofer.} It is completely appropriate to describe the effects of heavy new physics in terms of an effective theory with higher-order operators constructed from only SM degrees of freedom. In contrast if new physics is light enough to be produced in meson decays the theory needs to include the SM fields together with all light new physics degrees of freedom. In this talk I discuss three models of light new physics, in each of which the SM is extended by a single light new state, either an axion-like particle, a hidden photon or a right-handed neutrino.

\subsection{Axion-like particles}

\begin{figure} [h!]
  \begin{center}
    \includegraphics[scale=.45]{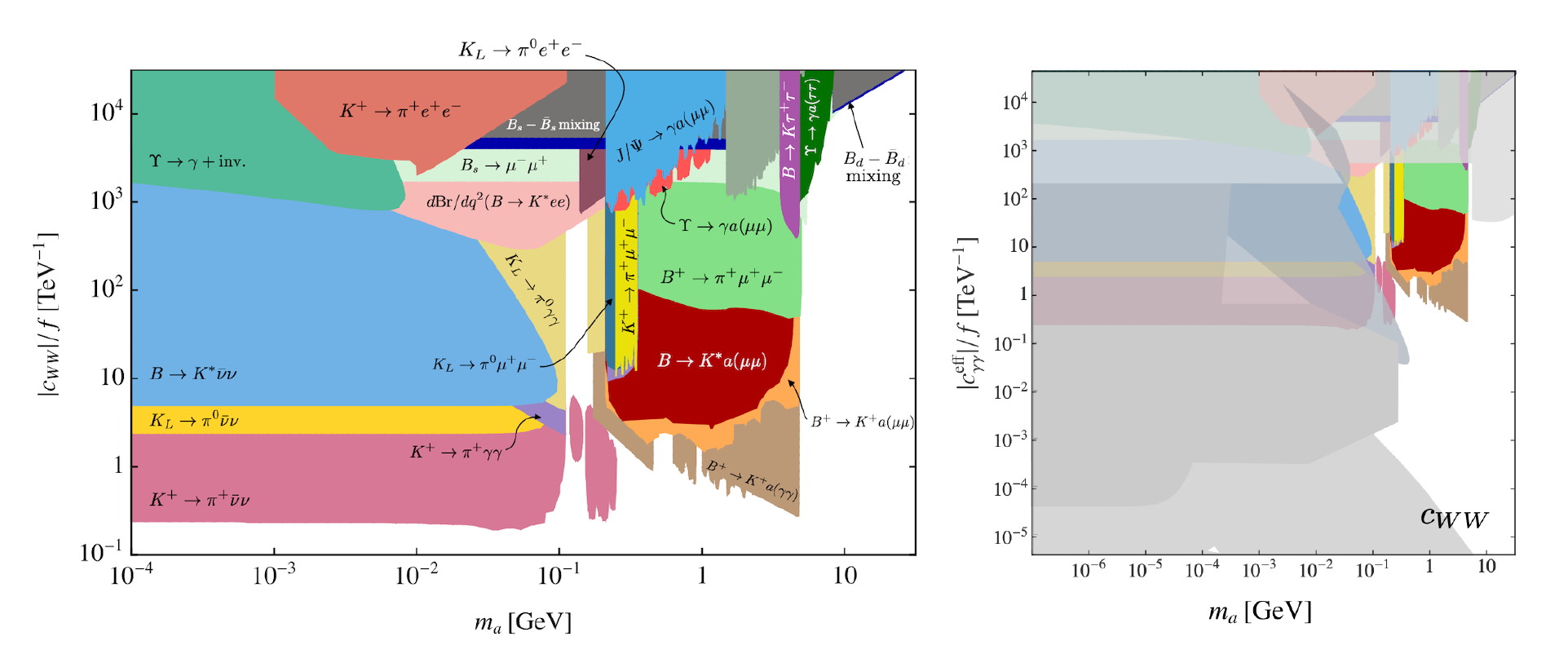}
    \caption{Comparison of constraints from various searches for a ALPs with a coupling to $SU(2)_L$ gauge bosons in the UV.  Figure from \cite{Bauer:2021mvw}. }
    \label{fig:ALPs}
  \end{center}
\end{figure}

Axion-like particles (ALPs) are pseudo Nambu Goldstone bosons that appear if global symmetries are spontaneously broken. If the underlying global symmetry was exact ALPs are massless, but if there is a source of explicit symmetry breaking an ALP mass proportional to the explicit breaking term is generated. The QCD pions are examples of such pseudo Nambu Goldstone bosons, whose masses vanish in the limit of zero quark masses for which the flavour symmetry of the SM is exact. As a result of the small first generation quark masses pions are about an order of magnitude lighter than non-pseudoscalar QCD resonances.  ALPs are interesting candidates for new physics because they could be low-mass harbinger of a ultraviolet (UV) theory at a scale that could be many orders of magnitude above the electroweak scale $\Lambda =4\pi f \gg v$.\\
ALPs interact with SM particles via shift-invariant operators apart from shift-symmetry breaking terms. We denote the pseudoscalar ALP as $a$ and write its couplings to SM fermions, gauge bosons and the Higgs boson as the most general Lagrangian to leading order in the new physics scale $f$ 
\begin{align}\label{Leff}
   {\cal L}_a
   &= \frac12 \left( \partial_\mu a\right)\!\left( \partial^\mu a\right) - \frac{m_{a}^2}{2}\,a^2
    + \frac{\partial^\mu a}{f}\,\sum_F\,\bar\psi_F\,c_F\,\gamma_\mu\,\psi_F 
    + c_\phi\,\frac{\partial^\mu a}{f}\, 
    \big( \phi^\dagger i \hspace{-0.6mm}\overleftrightarrow{D}\hspace{-1mm}_\mu\,\phi \big) \\
   &\quad\mbox{}+ c_{GG}\,\frac{\alpha_s}{4\pi}\,\frac{a}{f}\,G_{\mu\nu}^a\,\tilde G^{\mu\nu,a}
    + c_{WW}\,\frac{\alpha_2}{4\pi}\,\frac{a}{f}\,W_{\mu\nu}^A\,\tilde W^{\mu\nu,A}
    + c_{BB}\,\frac{\alpha_1}{4\pi}\,\frac{a}{f}\,B_{\mu\nu}\,\tilde B^{\mu\nu} \,.\notag
\end{align}
In the mass eigenbasis the ALP couplings to SM quarks are then given by matrices in flavour space $k_U$, $k_D$, $k_d$ and $k_{u}$ that are related to the matrices $c_F$ via the unitary rotations that diagonalise the Yukawa matrices, so that the ALP fermion couplings read
\begin{align}
{\cal L}_a^\text{fermion}
   = - \frac{ia}{2f}\,\sum_f\,\Big[ 
    &(m_{f_i}-m_{f_j}) \left[ k_f(\mu) + k_F(\mu) \right]_{ij} \bar f_i\,f_j\\
    &+ (m_{f_i}+m_{f_j}) \left[ k_f(\mu) - k_F(\mu) \right]_{ij} 
    \bar f_i\,\gamma_5 f_j \Big] \,.\notag
\end{align}
The precise structure of these couplings depends on the global symmetry of the UV theory. If the flavour structure of the ALP couplings to fermions in \eqref{Leff} allows for ALP mediated flavour-violating transitions at tree-level,  they set very strong constraints on the axion decay constant, e.g. the limit on $K\to \pi a$ set by NA62 corresponds to a constraint of $|k_D+k_d|_{12}/f < 1.2 \times 10^{-9}/\text{TeV}$~\cite{Bauer:2021mvw,Bauer:2021wjo}. 

Even if the ALP couplings are flavour-universal in the UV, running effects generate flavour-changing ALP couplings~\cite{Bauer:2021mvw}. Numerically one finds that below the electroweak scale and in the limit of zero quark masses apart from the top mass 
\begin{equation}\label{eq:FVveryshort}
\begin{aligned}
   \left[ k_D(m_t) \right]_{ij}^{\rm loop}
   \simeq V_{ti}^* V_{tj}\,\Big[&1.9\times 10^{-2}\,c_{tt}(\Lambda) 
    - 6.1\times 10^{-5}\,\tilde c_{GG}(\Lambda)\\
    & - 2.8\times 10^{-5}\,\tilde c_{WW}(\Lambda) 
    - 1.8\times 10^{-7}\,\tilde c_{BB}(\Lambda) \Big] \,. 
\end{aligned}
\end{equation}
where all couplings are defined at the UV scale $\Lambda$, $c_{tt}=(k_u)_{33}-(k_U)_{33}$ and the tilded coefficients are the physical ALP gauge boson couplings and include contributions from fermion couplings as well~\cite{Bauer:2020jbp}. In addition to running and matching effects the ALP mixes with the neutral QCD pseudo-Nambu Goldstone bosons upon matching to the chiral Lagrangian, which leads to flavour-changing couplings that are particularly sensitive to ALP interactions with gluons and quarks in the UV~\cite{Bauer:2021wjo}. Taking all contributions into account results in strong constraints on ALPs from searches in Kaons and B meson decays almost independent of the UV structure \eqref{Leff}. To demonstrate this the corresponding constraints are shown in the left panel of Figure~\ref{fig:ALPs} for an ALP with a coupling to $SU(2)_L$ gauge bosons in the UV.  In the right panel these constraints are compared to bounds from other experiments on the ALP coupling to photons~\cite{Bauer:2017ris}. Constraints from flavour observables, in particular searches for resonances in B meson decays can probe the 'gap' in parameter space for which ALPs are too short-lived to be detected in beam dump experiments but too light for collider searches to become sensitive.

\begin{figure} [h!]
  \begin{center}
    \includegraphics[scale=.43]{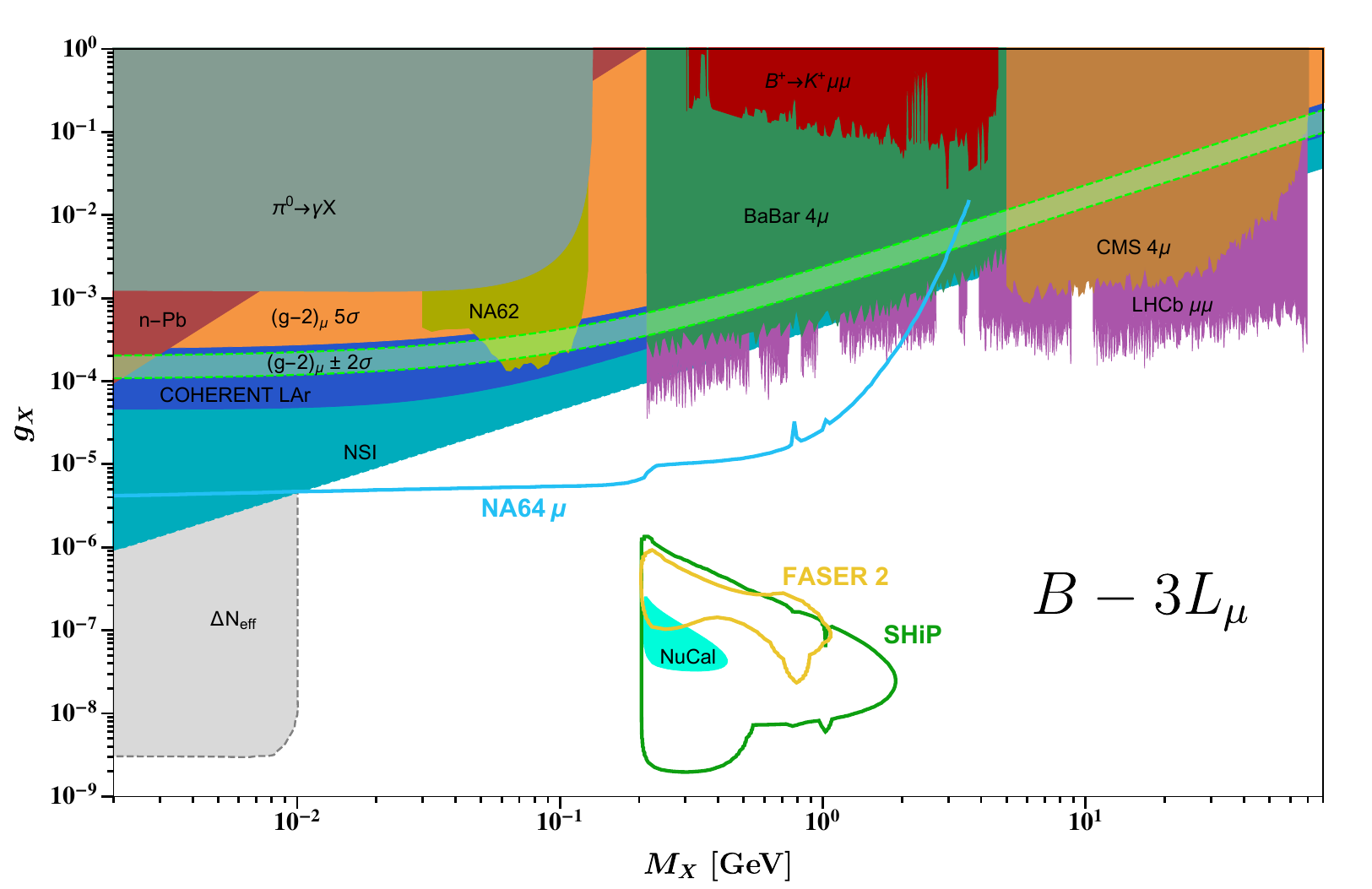}
    \caption{Comparison of constraints from various searches for a hidden photon with a $U(1)_{B-3L_\mu}$ symmetry group with the search for resonances in $B\to KX\to K \mu^+\mu^-$ decays. Figure from \cite{Bauer:2020itv}. }
    \label{fig:photons}
  \end{center}
\end{figure}

\section{Hidden Photons}

Hidden photons are spin-1 gauge bosons that interacts with the SM via kinetic mixing with the SM hypercharge gauge boson. In general there can also be SM particles charged under the $U(1)_X$ symmetry group that corresponds to the new spin-1 boson so that the Lagrangian includes the operators
\begin{align}
\mathcal{L}_X =\mathcal{L}_\text{SM} -\frac{1}{4} X_{\mu\nu} X^{\mu\nu}-\frac{\epsilon}{2} B_{\mu\nu} X^{\mu\nu} +\frac12 M_X^2 X_\mu X^\mu-g'j_\mu^Y B^\mu-g_Xj_\mu^X X^\mu\,,
\end{align}
where $B^{\mu\nu}$ is the hypercharge field strength tensor and $X^{\mu\nu}$ is the $U(1)_X$ field strength tensor, $g'$ and $g_X$ are the gauge couplings and $j_\mu^Y$ and $j_\mu^X$ are the associated currents. Even if absent at tree-level, the mixing parameter $\epsilon$ is generated at loop-level if any particles are charged under both hypercharge and $U(1)_X$. \\
Couplings of the hidden photon to SM quarks are diagonal in the quark mass basis. For $j^X_\mu=0$, the reason is that the hidden photon inherits the coupling structure of the SM hypercharge gauge boson through the kinetic mixing mechanism. Even at loop level flavour-violating decays can only be induced via mixing with the $Z$-boson component of the hypercharge gauge boson. Flavour-changing hidden photon couplings are therefore GIM-suppressed and need to vanish in the limit $M_X\to 0$. This argument doesn't change if SM quarks are charged under $U(1)_X$, since any way to assign charges to quarks for which all anomalies cancel and the CKM matrix can be reproduced without introducing additional particles enforce a universal coupling to SM quarks~\cite{Bauer:2020itv,Bauer:2022nwt}. One can write the effective flavour-violating couplings as 
\begin{align}\label{eq:currents}
\hspace{-.28cm}\mathcal{L}_X^\text{FCNC}&= \frac{M_X^2}{M_W^2}\bar d_j \gamma_\mu \left[g^L_{ij} P_L \!\!+g^R_{ij}P_R\right]\!  d_i X^\mu +\frac{g^\sigma_{ij}m_{d_j}}{2M_W^2}\bar d_j \sigma^{\mu\nu}\!\!\left[P_L+\frac{m_{d_i}}{m_{d_j}}P_R\right]\!d_i X_{\mu\nu}\,,
\end{align}
with $P_L, P_R$ projection operators and the couplings are given by
\begin{align}
g^L_{ij}=g_Xq_q \frac{\alpha}{8\pi s_w^2} V_{ti}V_{tj}^*f_1(x_t)\,,\qquad 
g^\sigma_{ij}&= g_X q_q\frac{\alpha}{8\pi s_w^2} V_{ti}V_{tj}^*f_2(x_t)\,,
\end{align}
and $g^R_{ij}=0$, where $q_q$ is the charge of the internal quarks, $x_t\equiv m_t^2/M_W^2$ and the loop functions $f_1(x_t)\approx 0.97$ and  $f_2(x_t)\approx -0.36$~\cite{Inami:1980fz}. 
The new gauge boson $X$ can be produced in flavour changing meson decays. For the example of the $B$-meson decay into Kaons and hidden photons one finds in the limit $m_d=0$ and $m_b\approx M_B$
\begin{align}
\Gamma(B &\to K X)\!=\!\frac{\lambda_K^{3/2}}{256\pi}\frac{M_B^2M_X^2}{M_W^4} \bigg[g_{32}^L  f_+(M_X^2) +g^\sigma_{32}f_T(M_X^2) \Big[1+\frac{M_K^2}{M_B^2}\Big]^{-1}\bigg]^2,
\end{align}
where $\lambda_K$ is a phase space factor and the vector and tensor currents in \eqref{eq:currents} induce decay widths proportional to the vector and tensor form factors \cite{Bailey:2015dka}.\\
As a result of this powerful suppression searches for resonances in flavour-violating decays don't lead to relevant constraints in the case of minimal hidden photon models. In Figure~\ref{fig:photons} the constraint from a resonance search in $B\to KX\to K \mu^+\mu^-$ decays is shown in dark red. The constraint has been calculated using the constraint from the LHCb search for a scalar resonance~\cite{LHCb:2016awg}. Other constrains from various different searches shown in Figure~\ref{fig:photons} are significantly stronger ~\cite{Bauer:2020itv,Bauer:2018onh,Ilten:2018crw}. For any other anomaly-free gauge group $U(1)_X$ this constraint is even less relevant. \\
Conversely one can consider searches for hidden photons in flavour violating decays as probes that can distinguish between minimal models of hidden photons and models that have non-conserved currents $j_\mu^X$. In these models resonance searches in flavour-violating meson decays are some of the most sensitive probes~\cite{Dror:2018wfl, Dror:2020fbh}.

\section{Heavy neutral leptons}

\begin{figure} [h!]
  \begin{center}
    \includegraphics[scale=.4]{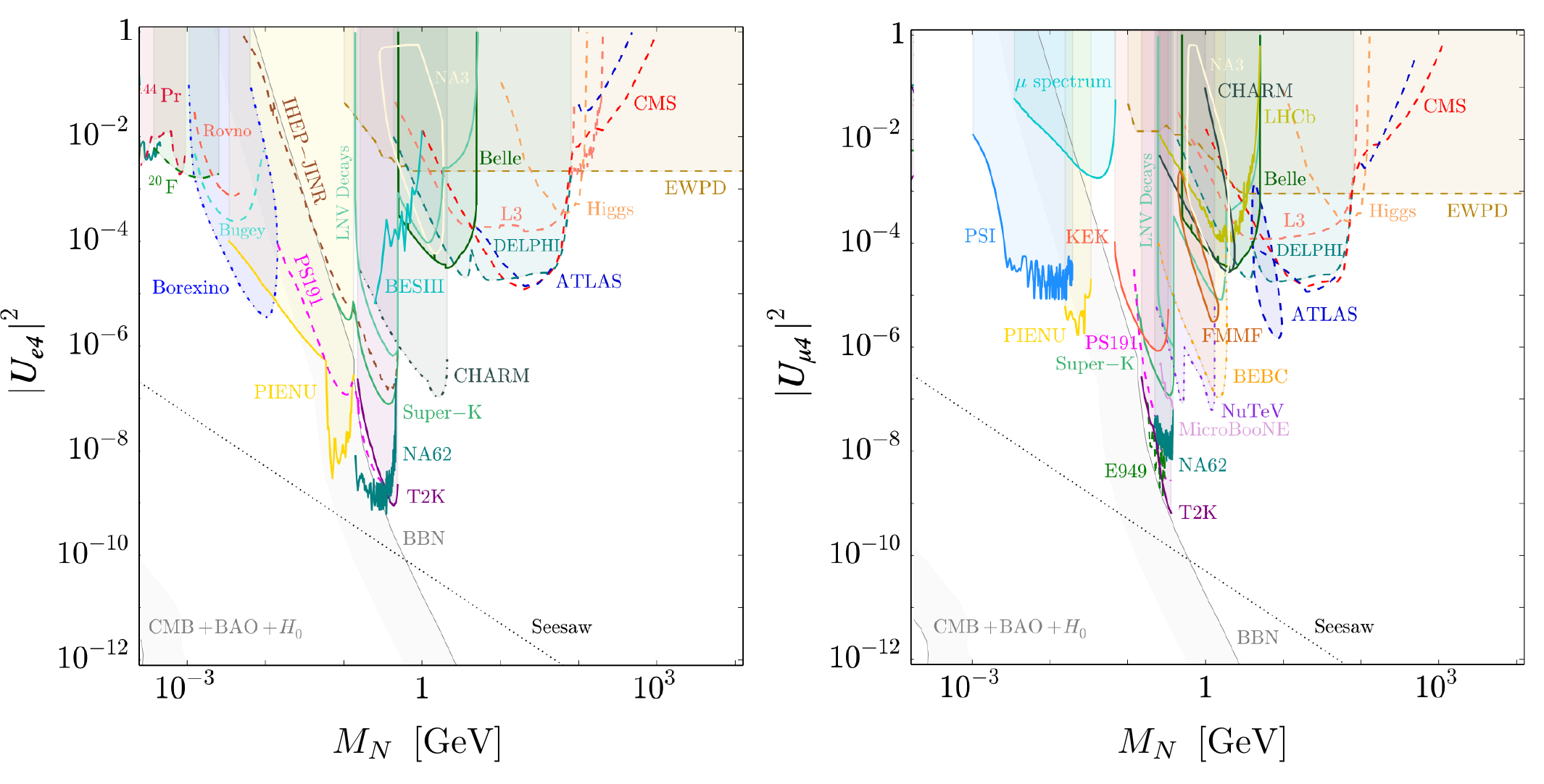}
    \caption{Comparison of constraints on the mixing angle of a single heavy neutral lepton of mass $M_N=1 \,\text{MeV} - 1\,\text{TeV} $ with electrons (left panel) and muons (right panel). The seesaw relation is indicated by the dotted diagonal line. Figure adapted from \cite{Dasgupta:2021ies}. }
    \label{fig:neutrinos}
  \end{center}
\end{figure}

One can write the Lagrangian of the SM extended by a three right-handed neutrino $N=(N_1,N_2,N_3)$
\begin{align}\label{eq:nLag}
\mathcal{L} = \mathcal{L}_{SM} + i \overline{N} \gamma^\mu \partial_\mu N - Y_\nu \overline{L} \tilde{\Phi} N - \frac{1}{2} \overline{N^c}M N + h.c.,
\end{align}
where $Y_\nu$ are the Yukawa couplings.\footnote{If one wants to consider the Lagrangian with just a single right-handed neutrino one simply sets $N=(N_1,0,0)$ in \eqref{eq:nLag}.} After electroweak symmetry breaking the neutrino masses are given by $M_N\approx M$ and 
\begin{align}
m_\nu =\frac{v^2}{2}\, Y_\nu M^{-1}Y_\nu^T = \theta M \theta^T\,,\qquad \text{with}\quad |\theta|^2=\frac{(vY_\nu)^2}{2}M^{-2}\sim \frac{m_\nu}{M_N}
\end{align}
where $\theta$ quantifies the mixing between the active neutrinos and the heavy mass eigenstates. In the generic see-saw mechanism the mixing angle is directly related to the light neutrino masses and therefore suppressed.
Since all flavour-violating interactions of the heavy neutral lepton are proportional to this mixing angle, flavour observables aren't very sensitive. Any interaction with the SM must vanish in the limit $m_\nu \to 0$ in which lepton number symmetry is restored. In models with multiple right-handed neutrinos additional lepton-number breaking parameters can break this relation and the mixing angle isn't necessarily suppressed~\cite{Abdullahi:2022jlv}.\\Some of the strongest bounds on the heavy neutrino mixing angle are obtained from resonance searches for HNLs in pion decays $\pi^+ \to e^+ N$ and $\pi^+ \to \mu^+ N$ by PIENU~\cite{PIENU:2017wbj, PIENU:2019usb} and Kaon decays $K^+\to e^+ N$  and $K^+\to \mu^+ N$ by NA62~\cite{NA62:2017qcd}. The corresponding constraints are shown in yellow (PIENU) and dark Green (NA62) in Figure~\ref{fig:neutrinos}. The bounds are comparable to the constraints from T2K (purple), a search for HNL decays also produced in Kaon decays~\cite{T2K:2019jwa}. It is interesting to note that these constraints are almost sensitive to the 'Seesaw' contour. \\
An alternative strategy is to search for lepton-number violating meson decays mediated by internal HNLs~\cite{Atre:2009rg}.  A search for $B^-\to \mu^+\mu^- \pi^+$ has been performed by LHCb~\cite{LHCb:2014osd} and for $D\to K\pi e^+e^+$ by BESIII~\cite{BESIII:2019oef}.  The constraint from BESIII is shown in the left panel in Figure~\ref{fig:neutrinos} in light blue and the constraint from LHCb is shown in the right panel in Figure~\ref{fig:neutrinos} in olive green.

\section{Summary}
Searches for light new physics in flavour violating transitions provide important limits on ALPs and HNLs. They are the most sensitive probes for parts of the parameter space that can't currently be probed by any other experiment. In contrast, minimal hidden photon models only allow for strongly suppressed flavour-changing interactions and any detection would be evidence of a more complicated new physics sector.

\bibliographystyle{amsplain}

\end{document}